\documentstyle[aps,preprint,epsfig,floats]{revtex}

\def\slash#1{#1\!\!\!/}
\def\eqref#1{Eq.\ (\ref{#1})}
\def\figref#1{Fig.\ \ref{#1}}

\begin{document}
\setcounter{page}{0}
\def\footnoterule{\kern-3pt \hrule width\hsize \kern3pt}
\tighten

\title{Opening the Crystalline Color Superconductivity Window}

\author{Adam~K.~Leibovich\footnote{Email address: {\tt adam@fnal.gov}}}

\address{Theory Group, Fermilab \\
P.O. Box 500, Batavia, IL 60510 \\
{~}}

\author{Krishna~Rajagopal\footnote{Email address: {\tt krishna@ctp.mit.edu}},
Eugene~Shuster\footnote{Email address: {\tt eugeneus@mit.edu}}}

\address{Center for Theoretical Physics \\
Massachusetts Institute of Technology \\
Cambridge, MA 02139 \\
{~}}

\date{MIT-CTP-3108,~ FERMILAB-Pub-01/041-T,~  hep-ph/0104073,~ April 4, 2001}
\maketitle

\thispagestyle{empty}

\begin{abstract}
Cold dense quark matter 
is in a crystalline color superconducting
phase wherever pairing occurs between species of quarks with
chemical potentials whose difference $\delta\mu$
lies within an appropriate window.  
If the interaction between
quarks is modeled as point-like, this window is rather
narrow.  
We show that when the interaction between quarks is modeled
as single-gluon exchange, the window widens by about a factor 
of ten at accessible densities and by much larger factors
at higher density.  This striking
enhancement reflects the increasingly
$(1+1)$-dimensional nature of the physics at weaker and weaker
coupling. Our results indicate that crystalline color
superconductivity is a generic feature of the phase diagram of 
cold dense quark matter, occurring wherever
one finds quark matter which is not in the color-flavor locked phase.
If it occurs within the cores of compact stars,
a crystalline color superconducting region
may provide a new locus for glitch phenomena.
\end{abstract}

\vfill\eject

\section{Introduction}

At asymptotic densities, the ground state of QCD with three
quarks with equal masses is expected to be the color-flavor locked (CFL) 
phase \cite{CFL,OtherCFL,ioffe,AlfordReview}.
This phase features a condensate of Cooper pairs of
quarks which includes $ud$, $us$, and $ds$ pairs. Quarks
of all colors and all flavors participate equally in the
pairing, and all excitations with quark quantum numbers are
gapped. 

The CFL phase persists for unequal quark masses, so long as the 
differences are not too large \cite{ABR2+1,SW2}.
In the absence of any interaction (and thus in the 
absence of CFL pairing) a quark mass difference
pushes the Fermi momenta for different
flavors apart, yielding different number densities for
different flavors.  In the CFL phase, however, the fact that
the pairing energy is maximized when $u$, $d$, and $s$ 
number densities are equal
enforces this equality \cite{rigidity}.  This means
that if one imagines increasing the strange quark
mass $m_s$, all quark number densities remain
equal until a first
order phase transition, at which CFL pairing is disrupted,
(some) quark number densities
spring free under the accumulated tension, and a less
symmetric state of quark matter is obtained \cite{rigidity}.
This
state of matter, which results when pairing between two species
of quarks persists even once their Fermi momenta differ,
is a crystalline color superconductor.

We can study crystalline color superconductivity more simply
by focusing just on
pairing between massless up and down quarks whose
Fermi momenta we attempt to push apart
by turning on a chemical
potential difference,
rather than a quark mass difference. 
That is, we introduce 
\begin{eqnarray}\label{mubardmu}
\mu_u&=&\bar\mu-\delta\mu\nonumber\\
\mu_d&=&\bar\mu+\delta\mu\ .
\end{eqnarray}
If $\delta\mu$ is nonzero but less than some $\delta\mu_1$,
the ground state 
is precisely that obtained for 
$\delta\mu =0$ \cite{Clogston,Bedaque,BowersLOFF}.  In this state,
red and green up and down quarks pair, yielding four quasiparticles
with superconducting gap $\Delta_0$ \cite{Barrois,BailinLove,ARW1,RappETC}.
Furthermore, the number density of red and green up quarks is the
same as that of red and green down quarks. 
As long as $\delta\mu$ is not too large, this BCS state
remains unchanged (and favored) 
because maintaining equal number densities, and
thus coincident Fermi surfaces, 
maximizes the pairing
and hence the gain in interaction energy.  
As $\delta\mu$ is increased, the BCS state remains the ground state of
the system only as long as its negative interaction
energy offsets the large positive free energy cost 
associated with forcing the Fermi seas to deviate from their
normal state distributions. In the weak coupling limit, in
which $\Delta_0/\bar\mu\ll 1$, the BCS state persists for
$\delta\mu<\delta\mu_1=\Delta_0/\sqrt{2}$ \cite{Clogston,BowersLOFF}.

These conclusions are the same (as long as $\Delta_0/\bar\mu\ll 1$)
whether the interaction between quarks is
modeled as a point-like four-fermion interaction or 
is approximated
by single-gluon 
exchange~\cite{ioffe}.
The latter 
analysis \cite{Son,PisarskiRischke,Hong,HMSW,SW3,rockefeller,Hsu2,BBS,ioffe,AlfordReview} is of quantitative validity at densities which are
so extremely high
that the QCD coupling $g(\mu)<1$ \cite{Shuster}.
Applying these asymptotic results at accessible densities 
nevertheless yields predictions for the magnitude
of the BCS gap $\Delta_0$, and thus
for $\delta\mu_1$, which are in qualitative agreement
with those made using phenomenologically normalized 
models with simpler point-like 
interactions \cite{SW3,ioffe,AlfordReview}.

Above the BCS state, in
a range $\delta\mu_1<\delta\mu<\delta\mu_2$,
the crystalline color
superconducting phase occurs.  We shall demonstrate in
this paper that applying asymptotic results at accessible
densities yields
a window $(\delta\mu_2-\delta\mu_1)$
that is more than a factor of ten wider than
in models with point-like interactions.
The crystalline color superconductivity window in parameter
space may therefore be much wider than previously thought,
making this phase a generic feature of the phase diagram
for cold dense quark matter.  The reason
for this qualitative increase in $\delta\mu_2$ can
be traced back to the fact that gluon exchange 
at weaker and weaker coupling 
is more and more dominated by forward-scattering, while
point-like interactions describe 
$s$-wave scattering.  What is perhaps surprising is that even
at quite {\it large} 
values of $g$, gluon exchange yields 
an order of magnitude increase in $\delta\mu_2-\delta\mu_1$.

The crystalline color superconducting state is the
analogue of a state
first explored by Larkin and Ovchinnikov~\cite{LO} 
and Fulde
and Ferrell~\cite{FF} in the context of electron superconductivity
in the presence of magnetic impurities. 
Translating LOFF's results to the case of interest, 
the authors of Ref. \cite{BowersLOFF}
found that for $\delta\mu\gtrsim\delta\mu_1$ it is
favorable to form a state in which the $u$ and $d$ Fermi
momenta are given by $\mu_u$ and $\mu_d$  as in the absence
of interactions, and are thus not equal, but pairing
nevertheless occurs.  Whereas in the BCS state, obtained
for $\delta\mu<\delta\mu_1$, pairing occurs between
quarks with equal and opposite momenta, when $\delta\mu\gtrsim\delta\mu_1$
it is favorable to form a condensate of Cooper pairs with nonzero
total momentum.  This is favored because 
pairing quarks with momenta which
are not equal and opposite gives rise to a region of phase space
where each quark in a Cooper pair can be
close to its own Fermi surface, even when the up and
down Fermi momenta differ, and such pairs can
be created at low cost in free energy.\footnote{LOFF
condensates have also recently been considered in two other contexts. 
In QCD with $\mu_u<0$, $\mu_d>0$ and $\mu_u=-\mu_d$, one has
equal Fermi momenta for $\bar u$ antiquarks and $d$ quarks,
BCS pairing occurs, and consequently a 
$\langle \bar u d\rangle$ condensate forms \cite{SonStephIsospin,Splittorff}.  
If $-\mu_u$ and $\mu_d$ differ,
and if the difference lies in the appropriate range, a LOFF
phase with a spatially varying $\langle \bar u d\rangle$
condensate results \cite{SonStephIsospin,Splittorff}.
Our conclusion that the LOFF window is much
wider than previously thought applies in this context also. 
Suitably isospin asymmetric nuclear matter may also admit LOFF pairing, as 
discussed recently in Ref. \cite{Sedrakian}.}
Condensates of this sort spontaneously break
translational and rotational invariance, leading to gaps which
vary periodically in a crystalline pattern.  If in some
shell within the quark matter core of a neutron star
(or within a strange quark star) the quark chemical potentials
are such that crystalline color superconductivity
arises, as we now see
occurs for a wide range of reasonable parameter
values,
rotational vortices may be pinned in this shell, making
it a locus for glitch formation \cite{BowersLOFF}. 
Rough estimates of the pinning force suggest that it 
is comparable to that for a rotational vortex pinned
in the inner crust of a conventional neutron star,
and thus may yield glitches of phenomenological interest \cite{BowersLOFF}.

As in Refs. \cite{BowersLOFF,BKRS}, we 
shall restrict our attention here to the 
simplest possible ``crystal'' structure, namely that in which
the condensate varies like a plane wave:
\begin{equation}\label{simplifiedcondensate}
\langle \psi({\bf x}) \psi({\bf x})\rangle \propto \Delta e^{2i{\bf q}
\cdot {\bf x}}\ .  
\end{equation} 
Wherever this condensate is favored over the homogeneous BCS
condensate and over the state with no pairing at all, we expect that
the true ground state of the system is a condensate which varies in
space with some more complicated spatial dependence.  The phonon
associated with the plane wave condensate (\ref{simplifiedcondensate})
has recently been analyzed \cite{CasalbuoniPhonon}. A full analysis of
all the phonons in the crystalline color superconducting phase must
await the determination of the actual crystal structure.

The authors of Refs. \cite{BowersLOFF,BKRS} studied crystalline
color superconductivity in a model in which
two flavors of quarks with chemical potentials (\ref{mubardmu})
interact via a point-like four-fermion interaction with
the quantum numbers of single gluon exchange. 
In the LOFF state,
each Cooper pair has total momentum $2{\bf q}$ with 
$|{\bf q}|\approx 1.2\delta\mu$ \cite{BowersLOFF}. The direction of
${\bf q}$ is chosen spontaneously. The LOFF
phase is characterized by a gap parameter $\Delta$ and
a diquark condensate, but not by an energy gap: the quasiparticle
dispersion relations vary with the direction of the momentum,
yielding gaps which range from zero up to a maximum of $\Delta$.
The condensate is dominated by those ``ring-shaped'' regions in momentum
space in which a quark pair with total momentum $2{\bf q}$
has both members of the pair within approximately $\Delta$ of their respective
Fermi surfaces.   

The gap equation which determines $\Delta$
was derived in Ref. \cite{BowersLOFF} using variational
methods, along the lines of Refs. \cite{FF,Takada2}.
It has since been rederived using now more familiar methods,
namely a diagrammatic derivation of the one-loop Schwinger-Dyson 
equation with a Nambu-Gorkov propagator modified to describe
the spatially varying condensate \cite{BKRS}.
This gap equation can then be used to show that crystalline
color superconductivity is favored over no $ud$ pairing
for $\delta\mu<\delta\mu_2$.  If quarks interact
via a weak point-like four-fermion interaction,
$\delta\mu_2\approx 0.754 \Delta_0$ \cite{LO,FF,Takada2,BowersLOFF}.
For a stronger four-fermion interaction, 
$\delta\mu_2$ tends to decrease,
closing the crystalline color superconductivity window \cite{BowersLOFF}. 

Crystalline color superconductivity is favored for 
$\delta\mu_1<\delta\mu<\delta\mu_2$. As $\delta\mu$ increases,
one finds a first order phase transition from the ordinary
BCS phase to the crystalline color superconducting phase
at $\delta\mu=\delta\mu_1$ and then a second order
phase transition at $\delta\mu=\delta\mu_2$ at which $\Delta$          
decreases to zero.
Because the condensation
energy in the LOFF phase is much smaller than that of the BCS condensate
at $\delta\mu=0$, the value of $\delta\mu_1$ is almost identical
to that at which the naive unpairing transition from the 
BCS state to the state with no pairing would occur if
one ignored the possibility of a LOFF phase, 
namely $\delta\mu_1=\Delta_0/\sqrt{2}$.  For all practical
purposes, therefore, the LOFF gap equation is not required in order
to determine $\delta\mu_1$. The LOFF gap equation
is used to determine $\delta\mu_2$
and the properties of the crystalline color
superconducting phase \cite{BowersLOFF}. 

In this paper, we generalize the diagrammatic analysis of
Ref. \cite{BKRS} to analyze the crystalline color
superconducting phase which occurs when quarks interact by
the exchange of a propagating
gluon, as is 
quantitatively valid at asymptotically high densities.
At weak coupling,
quark-quark scattering by single-gluon exchange is
dominated by forward scattering.  In most scatterings, 
the angular positions of the quarks on their 
respective Fermi surfaces do
not change much.  As a consequence, the weaker the coupling
the more the physics can be thought of as a sum of many
$(1+1)$-dimensional theories, with only rare large-angle scatterings
able to connect one direction in momentum space with 
others \cite{Hong}.
Suppose for
a moment that we were analyzing a truly $(1+1)$-dimensional
theory.
The momentum-space geometry of the LOFF state in one spatial dimension
is qualitatively different from that in three.  
Instead of Fermi surfaces, we would have
only  ``Fermi points'' at $\pm \mu_u$ and $\pm \mu_d$.
The only choice of $|{\bf q}|$ which allows pairing between
$u$ and $d$ quarks at their respective Fermi points
is $|{\bf q}|=\delta\mu$.
In $(3+1)$ dimensions, in contrast, $|{\bf q}|>\delta\mu$ is
favored because it allows LOFF pairing in ``ring-shaped''
regions of the Fermi surface, rather than 
just at antipodal points \cite{LO,FF,BowersLOFF}.
Also, in striking contrast to the $(3+1)$-dimensional case,
it has long been known that in a true $(1+1)$-dimensional 
theory with a point-like interaction between fermions,
$\delta\mu_2/\Delta_0\rightarrow\infty$ in the weak-interaction
limit \cite{LOFF1D}. 

We therefore expect that in $(3+1)$-dimensional
QCD with the interaction given by single-gluon exchange,
as $\bar\mu\rightarrow \infty$ and $g(\bar\mu)\rightarrow 0$ 
the $|{\bf q}|$ which characterizes the LOFF phase should
become closer and closer to $\delta\mu$ and $\delta\mu_2/\Delta_0$
should diverge.  We shall demonstrate both effects,  and
shall furthermore show that both are clearly in evidence
already at the rather large coupling $g=3.43$, corresponding
to $\bar\mu=400$ MeV using the conventions of Refs. \cite{SW3,Shuster}.
At this coupling, $\delta\mu_2/\Delta_0\approx 1.2$, 
meaning that $(\delta\mu_2-\delta\mu_1)\approx (1.2-1/\sqrt{2})\Delta_0$,
which is much
larger than $(0.754-1/\sqrt{2})\Delta_0$.
If we go to much higher densities, where the calculation
is under quantitative control, we find an even more striking
enhancement: when
$g=0.79$ we find $\delta\mu_2/\Delta_0 > 1000$!
We see that (relative to expectations based on experience
with point-like interactions)
the crystalline color superconductivity window
is wider by more than four orders of magnitude at this
weak coupling, and is about one order
of magnitude wider at accessible densities if weak-coupling results
are applied there.

In Section II, we present the gap equation derived 
from a Schwinger-Dyson equation as in 
Ref. \cite{BKRS}, but now assuming that quarks interact
via single-gluon exchange.  As our goal is the evaluation
of $\delta\mu_2$, the value of $\delta\mu$ at
which the crystalline color superconducting gap vanishes,
in Section III we take the $\Delta\rightarrow 0$ limit
of the gap equation.
Analysis of the resulting equation yields $\delta\mu_2$.  
In Section IV we look
ahead to implications of our central result that
the crystalline color superconductivity window is wider
by (an) order(s) of magnitude than previously thought.

\section{The Gap Equation for Crystalline Color 
Superconductivity at Weak Coupling}

In the crystalline color superconducting phase \cite{LO,FF,BowersLOFF,BKRS},
the condensate contains pairs
of $u$ and $d$ quarks with momenta such that the total momentum
of each Cooper pair is given by $2{\bf q}$, with the direction
of ${\bf q}$ chosen spontaneously.  Such a condensate varies
periodically in space with wavelength $\pi/|{\bf q}|$ as in
(\ref{simplifiedcondensate}).  Wherever there is an instability
towards
(\ref{simplifiedcondensate}), we expect that the true
ground state will be a crystalline condensate which varies in space
like a sum of several such plane waves with the same $|{\bf q}|$.
As in Refs. \cite{BowersLOFF,BKRS}, we focus here on finding
the region of $\delta\mu$ where an instability towards 
(\ref{simplifiedcondensate}) occurs, leaving the determination
of the favored crystal structure to future work.

We begin by reviewing the
field theoretic framework for the analysis of  crystalline color
superconductivity presented in Ref. \cite{BKRS}. 
In order to describe pairing between $u$ quarks with momentum 
${\bf p} + {\bf q}$ and $d$ quarks with momentum ${\bf -p} + {\bf q}$,
we must use a modified Nambu-Gorkov spinor defined as
\begin{equation} \label{Psi}
\Psi(p,q) = \left(\begin{array}{l} \psi_u(p+q) \\ \psi_d(p-q) \\
\bar\psi^T_d(-p+q) \\ \bar\psi^T_u(-p-q) \end{array}\right)\ .
\end{equation}
Note that by $q$ we mean the four-vector $(0,{\bf q})$. The
Cooper pairs have nonzero total momentum, and the ground state
condensate (\ref{simplifiedcondensate}) is static.
The momentum dependence of (\ref{Psi})
is motivated by the fact that in the presence of
a crystalline color superconducting condensate,
anomalous propagation does not only mean picking up or
losing two quarks from the condensate. It also means picking
up or losing momentum $2{\bf q}$.
The basis (\ref{Psi}) has been chosen so that the 
inverse fermion propagator in the crystalline
color superconducting phase is diagonal in $p$-space
and is given by
\begin{equation} \label{Sinv}
S^{-1}(p,q) = \left[\begin{array}{cccc} \slash{p}+\slash{q}+\mu_u \gamma_0
& 0 & -\bar{\bf \Delta}(p,-q) & 0 \\ 0 & \slash{p}-\slash{q}+\mu_d
\gamma_0 & 0 & \bar{\bf \Delta}(p,q) \\ -{\bf \Delta}(p,-q) & 0 &
(\slash{p}-\slash{q}-\mu_d \gamma_0)^T & 0 \\ 0 & {\bf \Delta}(p,q) 
& 0 & (\slash{p}+\slash{q}-\mu_u \gamma_0)^T \end{array}\right]\ ,
\end{equation}
where $\bar{\bf \Delta} = \gamma_0 {\bf \Delta}^{\dagger} \gamma_0$
and ${\bf \Delta}=C\gamma_5\epsilon^{\alpha\beta 3}$ 
is a matrix with color ($\alpha$, $\beta$)
and Dirac indices suppressed. Note that the condensate
is explicitly antisymmetric in flavor.
$2{\bf p}$ is the relative momentum
of the quarks in a given pair and is different for different
pairs. In the gap equation below, we shall integrate over $p_0$
and ${\bf p}$. As desired, the off-diagonal blocks describe
anomalous propagation in the presence of a condensate of
diquarks with momentum $2{\bf q}$. The choice
of basis we have made is analogous to that
introduced previously in the analysis of a crystalline
quark-antiquark condensate \cite{ChiralCrystal}.

We obtain the gap equation by solving
the one-loop Schwinger-Dyson equation given by
\begin{equation} \label{SDeq}
 S^{-1}(k,q)-S_0^{-1}(k,q) = -g^2 \int \frac{d^4p}{(2\pi)^4}
 \Gamma_\mu^A S(p,q)\Gamma_\nu^B D_{AB}^{\mu\nu}(k-p),
\end{equation}
where  $D_{AB}^{\mu\nu} = D^{\mu\nu} \delta_{AB}$ is the gluon
propagator, $S$ is the full quark propagator, whose inverse is 
given by (\ref{Sinv}), and $S_0$ is the fermion
propagator in the absence of interaction, given by $S$ with
${\bf \Delta}=0$. The
vertices are defined as follows: 
\begin{equation} \label{vertex}
\Gamma_\mu^A = \left(\begin{array}{cccc} \gamma_\mu\lambda^A/2 & 0 & 0
& 0 \\ 0 & \gamma_\mu\lambda^A/2 & 0 & 0 \\ 0 & 0 &
-(\gamma_\mu\lambda^A/2)^T & 0 \\ 0 & 0 & 0 &
-(\gamma_\mu\lambda^A/2)^T \end{array}\right) .
\end{equation}

In previous analyses \cite{BowersLOFF,BKRS}, a point-like
interaction was introduced by replacing $g^2 D^{\mu\nu}$ by
$g^{\mu\nu}$ times a constant.  Here, instead, we analyze
the interaction between quarks given by the exchange of a medium-modified
gluon and thus use a gluon propagator given by
\begin{equation} \label{gluonprop}
D_{\mu\nu}(p) = {P^T_{\mu\nu}\over p^2-G(p)} + {P^L_{\mu\nu}\over p^2-F(p)} -
\xi {p_\mu p_\nu \over p^4}\ ,
\end{equation}
where $\xi$ is the gauge parameter, 
$G(p)$ and $F(p)$ are functions of $p_0$ and $|\bf p|$, and the
projectors $P^{T,L}_{\mu\nu}$ are defined as follows:
\begin{equation} \label{PLT}
P^T_{ij} = \delta_{ij} - \hat{p_i}\hat{p_j}, \ P^T_{00}=P^T_{0i}=0, \
P^L_{\mu\nu} = -g_{\mu\nu} + {p_\mu p_\nu \over p^2} - P^T_{\mu\nu}.
\end{equation}
The functions $F$ and $G$ describe the effects of the medium
on the gluon propagator.  If we neglect the Meissner effect (that is,
if we neglect the modification of $F(p)$ and $G(p)$ due to the gap
$\Delta$ in the fermion propagator) then $F(p)$ describes
Thomas-Fermi screening and $G(p)$ describes Landau damping and they are
given in the hard dense loop (HDL) approximation by \cite{LeBellac}
\begin{eqnarray} \label{GF}
F(p) &=& m^2 {p^2\over|{\bf p}|^2} \left( 1 - {ip_0\over|{\bf p}|} Q_0
\left( {ip_0\over|{\bf p}|} \right) \right), 
\ \ \ \ \ \  {\rm with}\ Q_0(x) = {1\over2} \log
\left( {x+1\over x-1} \right), \nonumber\\
G(p) &=& {1\over2} m^2 {ip_0\over|{\bf p}|} \left[ \left( 1 - \left(
{ip_0\over|{\bf p}|} \right)^2 \right) Q_0 \left( {ip_0\over|{\bf p}|}
\right) + {ip_0\over|{\bf p}|} \right] \ ,
\end{eqnarray}
where $m^2 = g^2 \mu^2/\pi^2$ is the Debye mass for $N_f=2$.

The method of solving
the Schwinger-Dyson equation has been outlined in
Ref. \cite{BKRS}.
One obtains coupled integral equations for the gap parameters
$\Delta_1$, which describes particle-particle and hole-hole pairing,
and $\Delta_{2,3,4}$, which describe particle-antiparticle and
antiparticle-antiparticle pairing.  
Because only $\Delta_1$ describes pairing of quarks near their respective
Fermi surfaces, this is the gap parameter 
which describes the gap in the quasiparticle spectrum and
is therefore the gap parameter of physical interest.  For the same reason,
the gap equation is 
dominated by the terms containing $\Delta_1$. 
Upon dropping $\Delta_{2,3,4}$ and 
renaming $\Delta_1\rightarrow\Delta$,
the gap equation derived in Ref. \cite{BKRS} can be written
as:
%
\begin{eqnarray} \label{GapEq}
\Delta(k_0) & = & \frac{- i g^2}{3 \sin^2{\frac{\beta(k,k)}{2}}} 
  \int \frac{d^4p}{(2\pi)^4}
\frac{\Delta(p_0)}{(p_0+E_1)(p_0-E_2)} 
\nonumber \\
 &&\times \left[ \frac{C_F}{(k-p)^2 - F(k-p)} + \frac{C_G}{(k-p)^2 - G(k-p)} + 
    \frac{C_\xi \xi}{(k-p)^2}\right],
\end{eqnarray}
where
\begin{eqnarray} \label{Cdefs}
C_F & = & \cos^2{\frac{\beta(k,p)}{2}}
 \cos^2{\frac{\beta(p,k)}{2}} 
 - \cos^2{\frac{\beta(k,-p)}{2}} \cos^2{\frac{\beta(-p,k)}{2}} -
 \sin^2{\frac{\beta(k,k)}{2}} 
 \sin^2{\frac{\beta(p,p)}{2}} ,
\nonumber \\
C_G  &=& \frac{\cos{\beta(k,-p)} \cos{\beta(-p,k)}-\cos{\beta(k,p)}
\cos{\beta(p,k)} }{2} -  2
\sin^2{\frac{\beta(k,k)}{2}} \sin^2{\frac{\beta(p,p)}{2}}
\nonumber \\
 && -\, \cos{\alpha(k,p)}\left(\cos{\alpha(p,k)} \sin^2{\frac{\beta(k,-p)}{2}}
+ \cos{\alpha(-p,-k)} \sin^2{\frac{\beta(k,p)}{2}}\right)
\nonumber \\
 && -\, \cos{\alpha(-k,-p)} \left(\cos{\alpha(p,k)}
\sin^2{\frac{\beta(p,k)}{2}} + \cos{\alpha(-p,-k)}
\sin^2{\frac{\beta(-p,k)}{2}}\right) , 
\nonumber \\
C_\xi & = & \sin^2{\frac{\beta(k,p)}{2}}
\sin^2{\frac{\beta(p,k)}{2}} - \sin^2{\frac{\beta(k,k)}{2}}
\sin^2{\frac{\beta(p,p)}{2}} - \sin^2{\frac{\beta(k,-p)}{2}}
\sin^2{\frac{\beta(-p,k)}{2}}\nonumber \\
 && +\, \cos{\alpha(k,p)} \left(\cos{\alpha(p,k)}
\sin^2{\frac{\beta(k,-p)}{2}} + \cos{\alpha(-p,-k)}
\sin^2{\frac{\beta(k,p)}{2}}\right) \nonumber \\
 && +\, \cos{\alpha(-k,-p)} \left(\cos{\alpha(p,k)}
\sin^2{\frac{\beta(p,k)}{2}} + \cos{\alpha(-p,-k)}
\sin^2{\frac{\beta(-p,k)}{2}}\right) ,
\end{eqnarray}
and
\begin{eqnarray} \label{E12}
E_1({\bf p}) = & + \delta\mu + \frac{1}{2}
\left(|{\bf p}+{\bf q}|-|{\bf p}-{\bf q}|\right) + \frac{1}{2}
\sqrt{\left(|{\bf p}+{\bf q}|+|{\bf p}-{\bf q}|-2\bar{\mu}\right)^2 +
4 \Delta^2 \sin^2\frac{\beta(p,p)}{2}}, \nonumber \\
E_2({\bf p}) = & -\delta\mu - \frac{1}{2}
\left(|{\bf p}+{\bf q}|-|{\bf p}-{\bf q}|\right) + \frac{1}{2}
\sqrt{\left(|{\bf p}+{\bf q}|+|{\bf p}-{\bf q}|-2\bar{\mu}\right)^2 +
4 \Delta^2 \sin^2\frac{\beta(p,p)}{2}},
\end{eqnarray}
with
\begin{eqnarray} \label{Angles}
\cos{\alpha(k,p)} = \widehat{(k-q)}\cdot\widehat{(k-p)}\ , \nonumber \\
\cos{\beta(k,p)} = \widehat{(q+k)}\cdot\widehat{(q-p)} \ .
\end{eqnarray}
In the next section, we shall use this gap equation to obtain
$\delta\mu_2$, the upper boundary of the crystalline
color superconductivity window.

\section{The Zero-Gap Curve and the Calculation of $\delta\mu_2$}

Solving the full gap equation, \eqref{GapEq}, is numerically
challenging.  Fortunately, our task is simpler. As 
$\delta\mu \to \delta\mu_2$, the gap $\Delta \to 0$. 
Therefore, in order to determine $\delta\mu_2$ 
we need only analyze the
$\Delta \to 0$ limit of \eqref{GapEq}.  
Upon taking this limit, we shall find a ``zero-gap curve''
relating $\delta\mu$ and $|{\bf q}|$, as obtained for
a point-like interaction in Refs. \cite{FF,Takada2,BowersLOFF}.
The largest value of $\delta\mu$ on this curve is $\delta\mu_2$,
and the $|{\bf q}|$ at this point on the curve is the favored
value of $|{\bf q}|$ in the crystalline color superconducting
phase for $\delta\mu\rightarrow\delta\mu_2$.

To take the $\Delta \to 0$  limit, we first
divide both sides of \eqref{GapEq} 
by $\Delta(k_0)$.  We must be careful, however: simply cancelling
the $\Delta(p_0)/\Delta(k_0)$ which now occurs on the right-hand
side would 
yield an integral which diverges
in the ultraviolet.  This divergence is in fact regulated by
the $p_0$-dependence of $\Delta(p_0)$ itself. 
While the precise shape of $\Delta(p_0)$
remains unknown unless we solve the full 
gap equation (\ref{GapEq}), we assume $\Delta(p_0)$
has a similar $p_0$ dependence 
as in the BCS case \cite{SW3,Shuster}, where it is
relatively flat for $p_0<\mu$ and then decreases 
rapidly to zero as $p_0$ increases to several times $\mu$.
This rapid decrease  regulates the integral.  
This means that for $k_0 \ll \bar\mu$, we can approximate
$\Delta(p_0)/\Delta(k_0)$ by a step function which 
imposes an ultraviolet cutoff on the $p_0$ integral at
$p_0=\Lambda$, where $\Lambda$ is of order
several times $\bar\mu$.  
We shall investigate the dependence of
our results on the choice of $\Lambda$.     
After the factor $\Delta(p_0)/\Delta(k_0)$ has been replaced
by a cutoff, it is safe to simply set $\Delta$ to
zero everywhere else in the integrand.

After rotating to Euclidean space we obtain
\begin{eqnarray} \label{ZeroGapEq}
1 & = & \frac{g^2}{3 \sin^2{\frac{\beta(k,k)}{2}}} 
  \int^\Lambda \frac{d^4p}{(2\pi)^4}
\frac{1}{(p_0- i E_1)(p_0 + i E_2)} 
\nonumber \\
 &&\times \left[ \frac{C_F}{(k-p)^2 + F(k-p)} + \frac{C_G}{(k-p)^2 + G(k-p)} + 
    \frac{C_\xi \xi}{(k-p)^2}\right],
\end{eqnarray}
where the energies $E_{1,2}$ can be obtained from \eqref{E12} by setting
$\Delta$ equal to 0.
We also simplify the functions $F$ and $G$ to 
\begin{equation}
F(p) = m^2,\qquad\qquad G(p)=\frac{\pi}{4} m^2 \frac{p_0}{|{\bf p}|},
\end{equation}
valid for $p_0\ll|{\bf p}|\sim\bar\mu$ 
and to leading order in perturbation theory.
We shall work in the gauge with $\xi =0$, as this will allow
us to compare our results for $\delta\mu_2$ to values of $\Delta_0$
obtained in the same gauge in the analysis of the $\delta\mu=|{\bf q}|=0$
version of \eqref{GapEq} in Ref. \cite{SW3}.  As the 
gauge dependence in the calculation of  $\Delta_0$ decreases as 
$g$ is reduced below $\sim 1$ \cite{Shuster}, 
we expect that the same is true here for $\delta\mu_2$. We leave an analysis
of the $\xi$-dependence of $\delta\mu_2$ to the future.   
In evaluating the right-hand side of \eqref{ZeroGapEq}, we should
set $k_0\sim \Delta$, but instead choose $k_0=0$ for simplicity
as we expect $\Delta(k_0)$ to be almost constant 
for $k_0<\Delta$~\cite{SW3,Shuster}.
We must choose ${\bf k}$ in the region of momentum space
which dominates the pairing: for $\delta\mu=|{\bf q}|=0$, for example,
one takes $|{\bf k}|=\mu$ \cite{SW3,Shuster}.  
Now, we choose ${\bf k}$ on the ring in ${\bf k}$-space 
which describes pairs of quarks both of whose momenta (${\bf k}+{\bf q}$
and ${\bf k}-{\bf q}$)
lie on their respective noninteracting Fermi surfaces,
as this is where pairing is most important~\cite{LO,FF,BowersLOFF}.  
For $|{\bf q}|=\delta\mu\neq0$, as opposed
to $|{\bf q}|>\delta\mu\neq0$, the ring degenerates to
the point in momentum space with $|{\bf k}|=\bar\mu$ and 
${\bf k}$ antiparallel to ${\bf q}$.

We do the $p_0$ integral analytically.
There are six poles in the complex $p_0$
plane: two from the quark propagator 
located at $i E_1$ at $-i E_2$ and four from the gluon propagator 
located at
$\pm i \sqrt{|{\bf k} - {\bf p}|^2 + m^2}$, $-i (\pi m^2 \pm \sqrt{\pi^2
m^4+64 |{\bf k}-{\bf p}|^4})/(8 |{\bf k}-{\bf p}|)$.  
The poles from the gluon propagator have residues which 
are smaller than those from the quark propagator by factors
of $\bar\mu$.  We therefore keep only the   poles
at $i E_1$ and $-i E_2$.  
Furthermore, we can drop the $p_0^2$
pieces in the gluon propagators because at the $E_{1,2}$ 
poles they are negligible.
Upon doing the contour integral over $p_0$, we notice that
we only obtain a non-zero result if both $E_1$
and $E_2$ are positive. This defines
the ``pairing region'' $\mathcal P$ of 
Refs. \cite{BowersLOFF,BKRS}.  We get
\begin{equation} \label{FinalZeroGapEq}
1 = \frac{g^2}{3 \sin^2{\frac{\beta(k,k)}{2}}} 
  \int_{\mathcal P}^\Lambda \frac{p^2 dp\, d\Omega}{(2\pi)^3}
\frac{1}{(E_1 + E_2)} \left( \frac{C_F}{|{\bf k}-{\bf p}|^2 + F} + 
   \frac{C_G}{|{\bf k}- {\bf p}|^2 + G}\right),
\end{equation}
where the integral is taken over the pairing region $\mathcal P$ with
an upper cutoff on the $p$ integral of $\Lambda$.  The remaining
integrals are done numerically.

\begin{figure}[t]
\centerline{\epsfxsize 10cm \epsffile{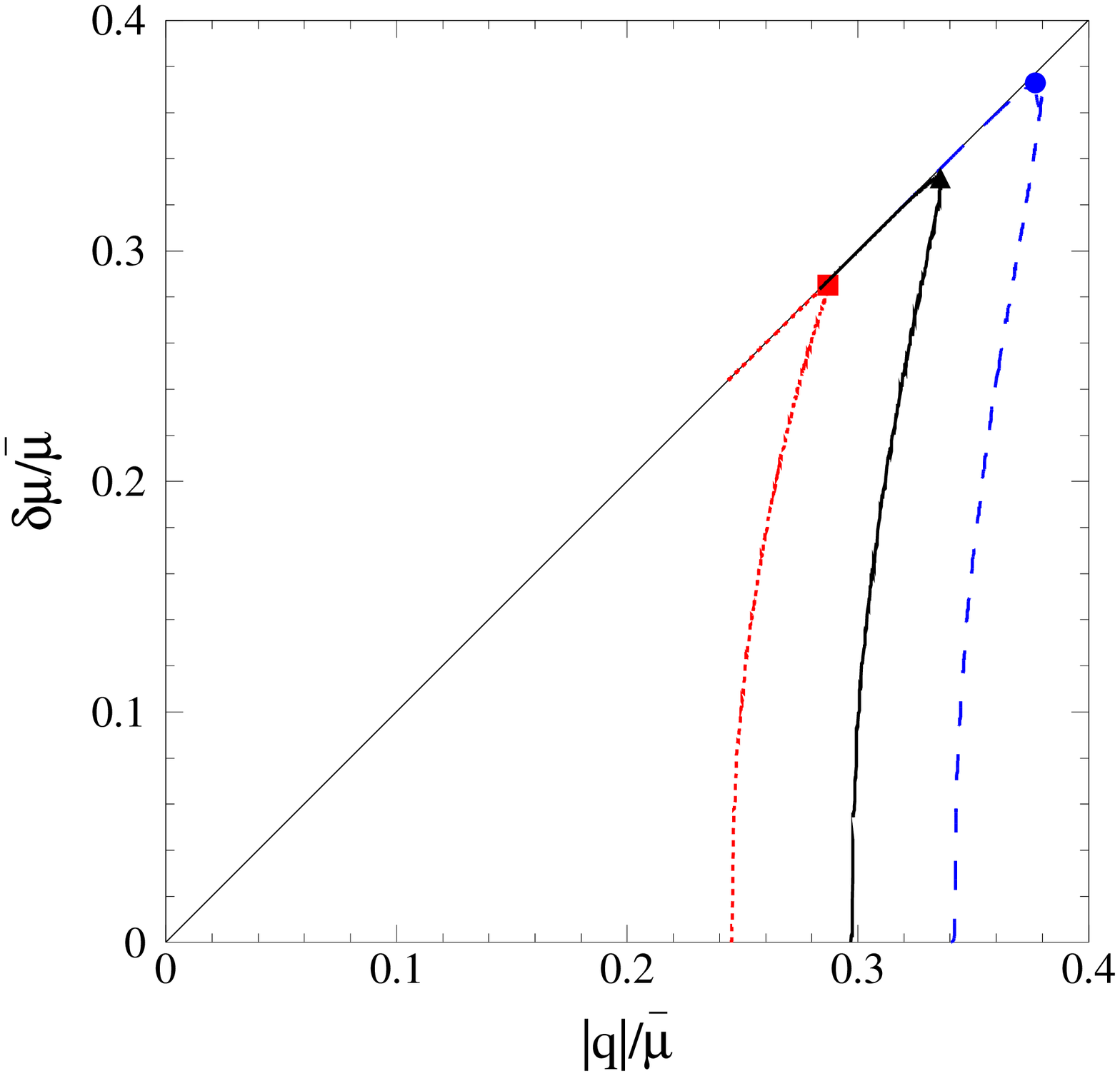}}
\centerline{\epsfxsize 10cm \epsffile{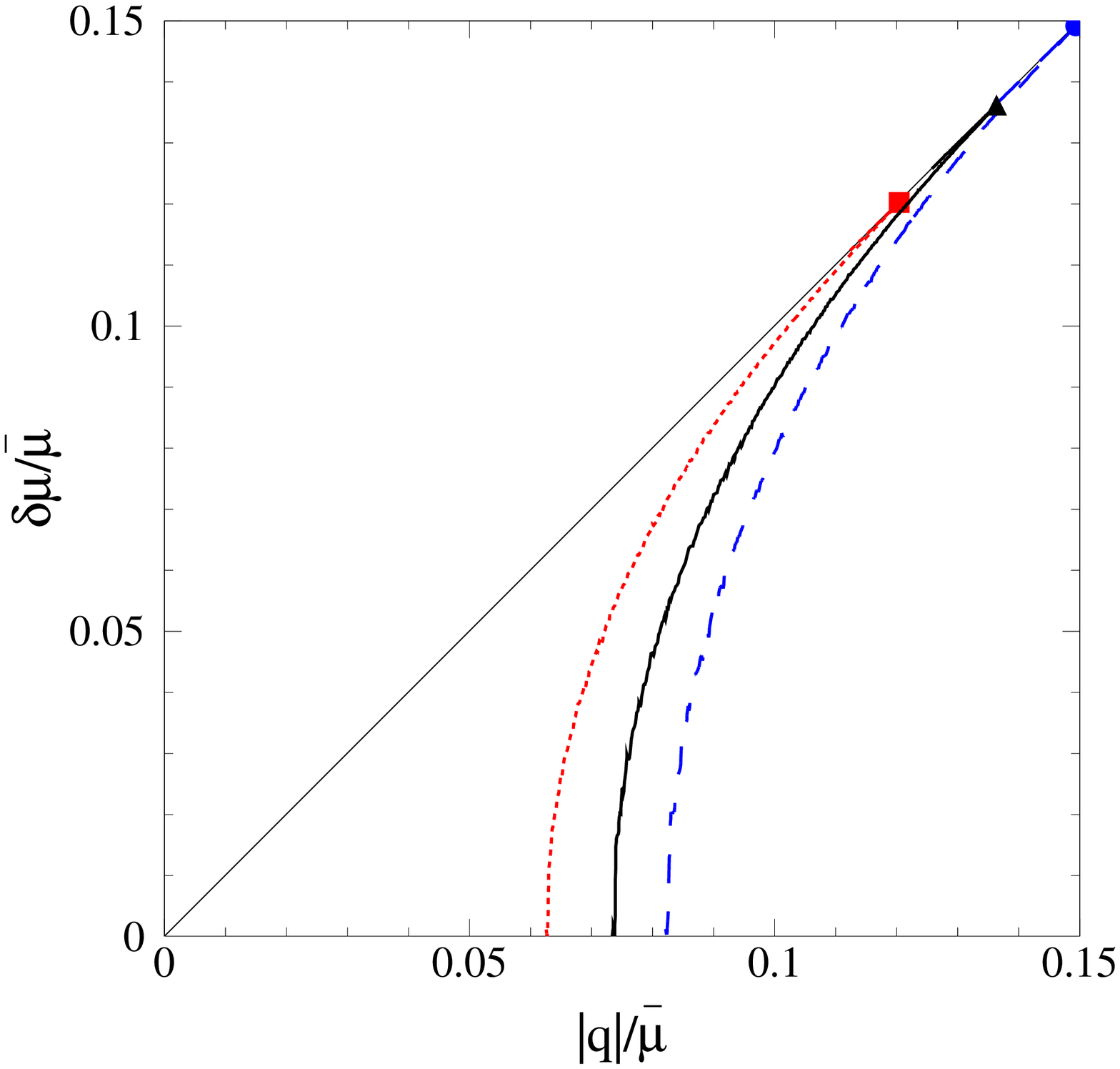}}
\caption{Zero-gap curves for different choices of the cutoff
$\Lambda$.  The upper panel is for $g=3.4328$, corresponding to
$\bar\mu = 400$~MeV, while the lower panel is for $g=2.2428$,
corresponding to $\bar\mu = 10^3$~MeV.  The three curves in each panel
correspond to $\Lambda/\bar\mu=2.0$ (dotted red), 
2.5 (solid black), and 3.0 (dashed blue).  
The square, triangle, and circle mark the locations of
$\delta\mu_2$ for each of the three curves.
}
\label{VaryLambdaFig}
\end{figure}
Upon doing the integrals,
\eqref{FinalZeroGapEq} becomes an equation relating $|{\bf q}|$ and
$\delta\mu$.  We explore the $(|{\bf q}|,\delta\mu)$ plane, evaluating
the right-hand side of \eqref{FinalZeroGapEq} at each point, and
in so doing map out the ``zero-gap curves'' along which 
\eqref{FinalZeroGapEq} is satisfied.  Several zero-gap curves are shown in
\figref{VaryLambdaFig}, in order to
exhibit the dependence of the curve
on the coupling $g$ and on the parameter
$\Lambda$.  The top panel is for 
$g=3.4328$: if we use one-loop running with $\Lambda_{\rm QCD}=200$ MeV
and $N_f=2$, this corresponds to $g(\bar\mu)$ for $\bar\mu=400$ MeV.  
The lower panel is for $g=2.2528$ corresponding in the same sense to
$\bar\mu=10^3$ MeV.
The three curves in each plot
are drawn with  $\Lambda/\bar\mu = 2.0$, 2.5,  
and 3.0.  
The choice of $\Lambda$ makes some
difference to the scale of the zero-gap curves, but does
not change the qualitative behavior.  
We are only
interested in the qualitative behavior, since the couplings $g$ which
are appropriate at accessible densities are much too large for
the analysis to be under quantitative control anyway.
In subsequent calculations, we shall take $\Lambda/\bar\mu =2.5$ 
as our canonical choice.
Removing the artificially
introduced parameter $\Lambda$ would require solving the 
full gap equation (\ref{GapEq}).

We only show the zero-gap curve for $|{\bf q}|\geq \delta\mu$,
because the favored value of $|{\bf q}|$ must lie
in this region \cite{LO,FF,Takada2,BowersLOFF}.
Each zero-gap curve in \figref{VaryLambdaFig} begins somewhere
on the line $\delta\mu=|{\bf q}|$, moves up and to the right
following the line $\delta\mu=|{\bf q}|$ very closely,
and then turns around sharply and heads downward, eventually
reaching the $\delta\mu=0$
axis.  This behavior is most clearly seen in the upper panel.
The smaller the value of $g$, the more sharply the curve 
turns around and the more closely the downward-going curve
hugs the $\delta\mu=|{\bf q}|$ line. 
Solutions to the full gap equation (\ref{GapEq})
with $\Delta\neq 0$ occur in the region bounded by 
the zero-gap curve, between it and the $\delta\mu=|{\bf q}|$ line.
As $g$
is decreased, this region
becomes a sharper and sharper sliver, squeezed closer
and closer to the $\delta\mu=|{\bf q}|$ line.  
Since in $(1+1)$ dimensions $|{\bf q}|=\delta\mu$ is favored, 
the change in the zero-gap curves from the upper panel
to the lower 
vividly 
demonstrates how the physics embodied in the gap equation (\ref{GapEq})
becomes effectively $(1+1)$-dimensional as $g$ is reduced.

\begin{figure}[t]
\centerline{\epsfxsize 18cm \epsffile{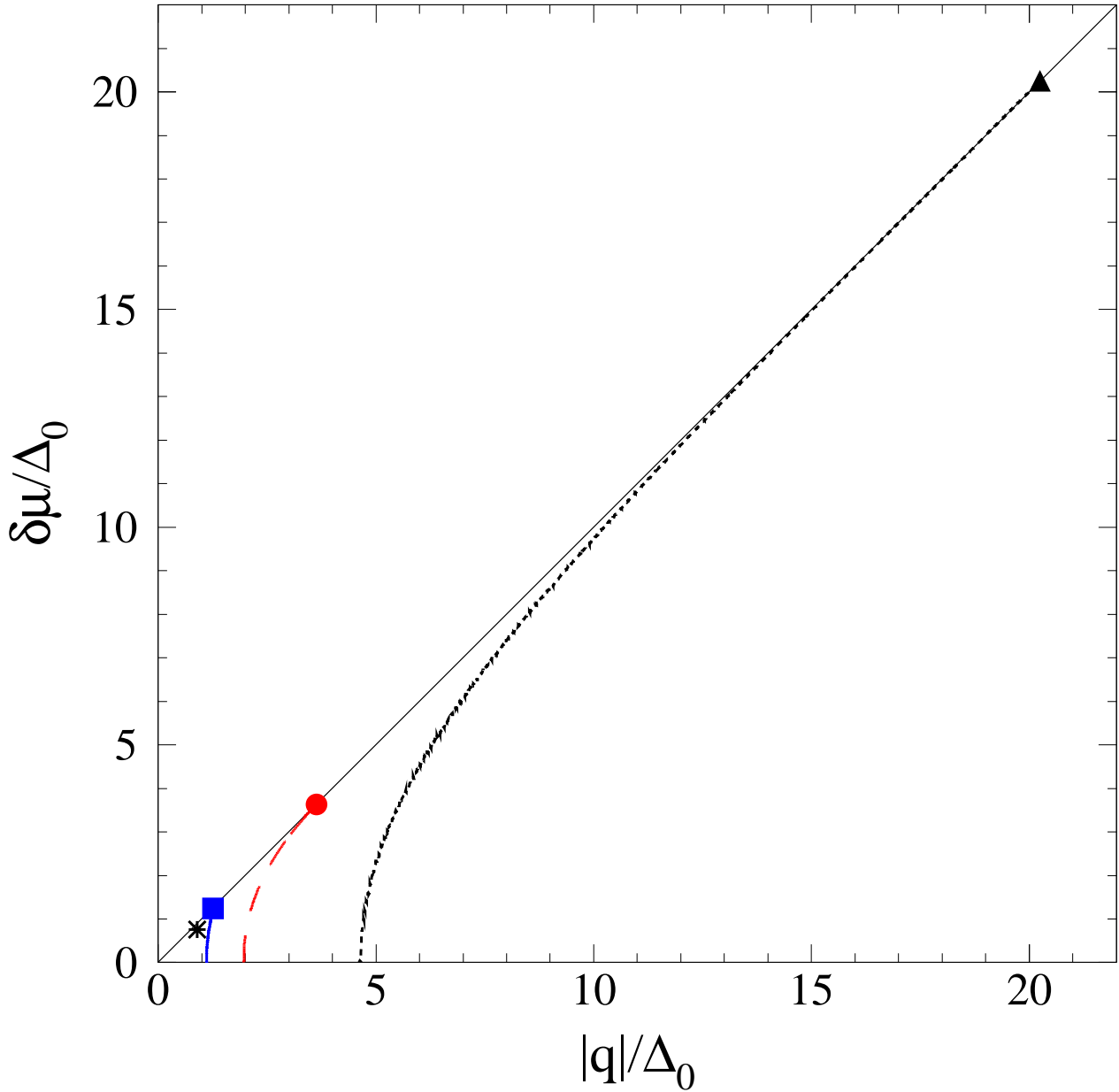}}
\caption{Zero-gap curves for different choices of $\bar\mu$.  Each
curve is normalized to the BCS gap $\Delta_0$ for the corresponding
$\bar\mu$. The three curves are 
$\bar\mu = 400$ MeV (solid blue), $10^3$
MeV (dashed red), and $10^4$ MeV (dotted black).  Again, $\delta\mu_2$
is marked for each curve.  For 
comparison, the star is the position of
$\delta\mu_2$ for a weak point-like interaction.}
\label{VaryMuFig}
\end{figure}
The largest value of $\delta\mu$ on the zero-gap curve
is $\delta\mu_2$. For $\delta\mu<\delta\mu_2$, a range of
values  of $|{\bf q}|$
can be found for which a crystalline color superconducting
phase with $\Delta\neq 0$ exists. The $|{\bf q}|$ within this
range which results in a phase with the lowest free energy
is favored, and the corresponding $\Delta$ obtained
by solving \eqref{GapEq} 
at that $|{\bf q}|$ is favored.  At $\delta\mu_2$, however,
$\Delta\rightarrow 0$. For each curve in  \figref{VaryLambdaFig},
the location of $\delta\mu_2$ is marked.  In all
cases,  $|{\bf q}|$ at $\delta\mu_2$ is very close to $\delta\mu_2$.


We see from \figref{VaryLambdaFig}
that $\delta\mu_2/\bar\mu$ decreases as $g$ decreases,
but this is not the important comparison.  Recall
that the crystalline color superconducting phase 
occurs within a window $\delta\mu_1<\delta\mu<\delta\mu_2$
where $\delta\mu_1=\Delta_0/\sqrt{2}$ at weak coupling,
with $\Delta_0$ the BCS gap obtained at $\delta\mu=0$.
To evaluate the width of the window, we must therefore also
compare $\delta\mu_2$ to $\Delta_0$. We obtain $\Delta_0$
from Ref. \cite{SW3}, and in  \figref{VaryMuFig} we 
plot the zero-gap curves for 
$\bar\mu$ equal to
$400$~MeV, $10^3$~MeV, 
and $10^4$~MeV, in each case scaled by
the corresponding BCS gap $\Delta_0$.  
In contrast, the star marks the location
of $\delta\mu_2$ obtained
in a theory with a weak point-like
interaction: $\delta\mu_2=0.754\Delta_0$ and $|{\bf q}|=0.906\Delta_0$.

\begin {table}[t]
\begin {center}
\begin {tabular}{cccccc}
$\bar\mu$ (MeV) & $g$ & $\delta\mu_2/\bar\mu$ & $|{\bf q}|/\delta\mu_2$ & 
$\Delta_0$ (MeV) & $\delta\mu_2/\Delta_0$ \\
\hline
400 & 3.4328 & 0.3317 & 1.012 & 107 & 1.24 \\
$10^3$ & 2.2528 & 0.1362 & 1.001 & 37.5 & 3.63 \\
$10^4$ & 1.4450 & 0.0241 & 1.000 & 11.9 & 20.3 \\
$10^5$ & 1.1464 & 5.23 $\times 10^{-3}$ & 1.0000 & 8.0 & 65.3 \\
$10^6$ & 0.9793 & 1.71 $\times 10^{-3}$ & 1.0000 & 7.63 & 225 \\
$10^7$ & 0.8689 & 5.84 $\times 10^{-4}$ & 1.0000 & 9.16 & 638 \\
$10^8$ & 0.7890 & 1.63 $\times 10^{-4}$ & 1.0000 & 13.0 & 1254 \\
\end {tabular}
\end {center}
\caption{Positions of $\delta\mu_2$ and corresponding $|{\bf q}|$
for seven values of $g$: 
if we use one-loop running with $\Lambda_{\rm QCD}=200$~MeV,
these correspond to $g(\bar\mu)$ for the 
values of $\bar\mu$ shown. For each $g$, we give $\delta\mu_2/\bar\mu$
and $|{\bf q}|/\delta\mu_2$ as determined from the zero-gap curves
we calculate.  To obtain $\delta\mu_2/\Delta_0$, we take $\Delta_0$
from Ref. \protect\cite{SW3}.}
\label{QvsMuTable}
\end {table}
The effect of the single-gluon exchange interaction, then,
is to increase $\delta\mu_2/\Delta_0$ dramatically.
Because this interaction favors forward scattering
at weak coupling, it causes $\delta\mu_2$ 
to approach the line $|{\bf q}| = \delta\mu$ and 
$\delta\mu_2/\Delta_0$ to diverge, as occurs in 
$(1+1)$ dimensions where there are only ``Fermi points''.
In Table \ref{QvsMuTable}, 
we show the
$\delta\mu_2/\Delta_0$ values for different values of $\bar\mu$, with
all calculations done for $\Lambda/\bar\mu=2.5$.
Already at $g=3.4328$, corresponding to $\bar\mu =400$~MeV,
$\delta\mu_2$ is within 1\% of
the $|{\bf q}| = \delta\mu$ line and $\delta\mu_2=1.24\Delta_0$.
Although this result will depend somewhat on $\Lambda$, it
means that the crystalline color superconducting window
is about ten times wider than for a point-like interaction.
Once $g=1.445$, corresponding to $\bar\mu =10^4$~MeV,
$\delta\mu_2$ is closer than one part in a thousand to
the $|{\bf q}| = \delta\mu$ line and $\delta\mu_2=20.3\Delta_0$,
meaning that the crystalline color superconducting window
is about four hundred times wider than for a point-like interaction.


\section{Implications and Future Work}

We have found that $\delta\mu_2/\Delta_0$ 
diverges in QCD as the weak-coupling, high-density limit
is taken.  
Applying results valid at asymptotically
high densities to those of interest in
compact stars, namely $\bar\mu\sim 400$ MeV, we find
that even here the crystalline color superconductivity
window is an order of magnitude wider than that obtained
previously 
upon approximating the interaction between quarks as point-like.

This discovery has significant implications for the QCD
phase diagram and may have significant
implications for compact stars.  At high enough baryon 
density
the CFL phase in which all quarks pair to form
a spatially 
uniform BCS condensate is favored. Suppose that as
the density is lowered the nonzero strange quark
mass induces the formation
of some less symmetrically paired quark matter
before the density is lowered so much that baryonic
matter is obtained. In this less symmetric quark matter,
some quarks may yet form a BCS condensate.  Those which
do not, however, will have differing Fermi momenta.
These will form a crystalline color superconducting
phase if the differences between their Fermi momenta
lie within the appropriate window.
In QCD, the interaction between quarks
is forward-scattering dominated and the 
crystalline
color superconductivity window is consequently
wide open. This phase is therefore generic,
occurring almost anywhere there 
are some quarks which cannot form BCS pairs.  
Evaluating
the critical temperature $T_c$ above which the crystalline
condensate melts requires solving the 
nonzero temperature gap equation obtained
from \eqref{GapEq} as described in Ref. \cite{BKRS}, 
but we expect that all but the very youngest compact
stars are colder than $T_c$.  This suggests that wherever
quark matter which is not in the CFL phase occurs 
within a compact star, rotational vortices may be pinned
resulting in the generation of glitches as the star
spins down.

Solidifying the implications of our results requires
further work in several directions.  First, we must
confirm that pushing Fermi surfaces apart via
quark mass differences has the same effect as pushing
them apart via a $\delta\mu$ introduced by hand.
Second, we must extend the analysis to the three
flavor theory of interest.  Third, we need to 
re-evaluate $\delta\mu_2$ by comparing
the crystalline color superconducting phase 
to a BCS phase in which spin-1 pairing between quarks
of the same flavor is allowed \cite{Schaefer1Flav}.  
The results of 
Ref. \cite{BowersLOFF} suggest that this can be neglected,
but confirming this in the present case 
requires solving
the full gap equation (\ref{GapEq}).  And, fourth,
before evaluating the pinning force on a rotational
vortex and making predictions for glitch phenomena,
we need to understand which crystal structure is favored.

\acknowledgments

We are grateful to J. Bowers, J. Kundu, D. Son, and
F. Wilczek for helpful conversations.
The work of AL is supported in part by the U.S. Department of Energy
(DOE) under grant number DE-AC02-76CH03000.  The work of KR and ES
is supported in part by the DOE under cooperative research agreement
DE-FC02-94ER40818. The work of KR is supported in part by a DOE OJI
grant and by the Alfred P. Sloan Foundation.

\end{document}